\documentclass[final,3p,times,twocolumn]{elsarticle}

\usepackage{graphicx}

\usepackage{amssymb}

\usepackage{lineno}
\usepackage[section]{placeins}

\usepackage{amsmath}
\usepackage{bm}
\usepackage{array}\allowdisplaybreaks[1]
\usepackage{braket}
\usepackage{slashed}
\usepackage{siunitx}
\usepackage{ulem,xpatch}
\usepackage{subfigure}

\usepackage{comment}
\usepackage{soul}
\usepackage{xcolor}
\usepackage{hyperref}
\usepackage{xfrac}
\hypersetup{colorlinks, linkcolor = [rgb]{1,0,0}, citecolor = [rgb]{1,0,0}, urlcolor = [rgb]{1,0,0}}

\newcommand{\Tr}{\mathrm {Tr}}

\newcommand{\nn}{\nonumber}

\def\be{\begin{equation}}
\def\ee{\end{equation}}
\def\bea{\begin{eqnarray}}
\def\eea{\end{eqnarray}}

\begin{document}

\title{Constraints on charm-anticharm asymmetry in the nucleon from lattice QCD}


\author[JLAB]{Raza~Sabbir~Sufian}
\author[JLAB]{Tianbo~Liu}
\author[GWU,UMD]{Andrei~Alexandru}
\author[SLAC]{Stanley~J.~Brodsky}
\author[UCR]{Guy~F.~de~T\'eramond}
\author[HU]{Hans~G\"unter~Dosch}
\author[UKY]{Terrence~Draper}
\author[UKY]{Keh-Fei~Liu}
\author[ITP,SFP,ICTPA]{Yi-Bo~Yang}

\address[JLAB]{Thomas Jefferson National Accelerator Facility, Newport News, VA 23606, USA}
\address[GWU]{Department  of  Physics,  The  George  Washington  University,  Washington,  DC  20052,  USA}
\address[UMD]{Department of Physics, University of Maryland, College Park, MD 20742, USA}
\address[SLAC]{SLAC  National  Accelerator  Laboratory,  Stanford  University,  Stanford,  CA  94309,  USA}
\address[UCR]{Laboratorio de F\'isica Te\'orica y Computacional, Universidad de Costa Rica, 11501 San Jos\'e, Costa Rica}
\address[HU]{Institut f\"ur Theoretische Physik der Universit\"at, D-69120 Heidelberg, Germany}
\address[UKY]{Department of Physics and Astronomy, University of Kentucky, Lexington, Kentucky 40506, USA}
\address[ITP]{CAS Key Laboratory of Theoretical Physics, Institute of Theoretical Physics, Chinese Academy of Sciences, Beijing 100190, China}
\address[SFP]{School of Fundamental Physics and Mathematical Sciences, Hangzhou Institute for Advanced Study, UCAS, Hangzhou 310024, China}
\address[ICTPA]{International Centre for Theoretical Physics Asia-Pacific, Beijing/Hangzhou, China}

\begin{abstract}
We present the first lattice QCD calculation of the charm quark contribution to the nucleon electromagnetic form factors $G^c_{E,M}(Q^2)$ in the momentum transfer range $0\leq Q^2 \leq 1.4$~$\rm GeV^2$. The quark mass dependence, finite lattice spacing and volume corrections are taken into account simultaneously based on the calculation on three gauge ensembles including one at the physical pion mass. The nonzero value of the charm magnetic moment $\mu^c_M=-0.00127(38)_{\rm stat}(5)_{\rm sys}$, as well as the Pauli form factor, reflects a nontrivial role of the charm sea in the nucleon spin structure. The nonzero $G^c_{E}(Q^2)$ indicates  the existence of  a nonvanishing asymmetric charm-anticharm sea in the nucleon. Performing a nonperturbative analysis based on holographic QCD and the generalized Veneziano model, we study  the constraints  on the $[c(x)-\bar{c}(x)]$ distribution from the  lattice QCD results presented here. Our results  provide complementary information and motivation for more detailed studies of physical observables that are sensitive to intrinsic charm and for future global analyses of parton distributions including  asymmetric charm-anticharm distribution.
\end{abstract}

\begin{keyword}
Intrinsic charm \sep Form factor \sep Parton distributions \sep Lattice QCD, Light-front holographic QCD
\sep
JLAB-THY-20-3155, SLAC-PUB-17515
\end{keyword}

\maketitle
\section{Introduction}\label{sec:intro}
The charm-anticharm sea in the nucleon has received great interest in nuclear and particle physics for its particular significance in understanding high energy reactions associated with charm production. Quantum Chromodynamics (QCD), the underlying theory of the strong interaction, allows heavy quarks in the nucleon-sea to have both perturbative ``extrinsic" and nonperturbative ``intrinsic" origins.  The extrinsic sea arises  from gluon splitting triggered by a probe in the reaction. It can be calculated order-by-order in perturbation theory if the probe is hard. The intrinsic sea is encoded in the nucleon wave functions. 

The existence of nonperturbative intrinsic charm (IC) was originally proposed in the BHPS model~\cite{Brodsky:1980pb} and in the subsequent calculations~\cite{Brodsky:1984nx,Harris:1995jx,Franz:2000ee} following the  original proposal~\cite{Brodsky:1980pb}. Proper knowledge of the existence of IC and an estimate of its magnitude will elucidate  some fundamental aspects of nonperturbative QCD. Therefore, the main goal of this article is to investigate the existence of nonzero ``intrinsic" charm of nonperturbative origin in the nucleon. In the case of light-front (LF) Hamiltonian theory, the intrinsic heavy quarks of the proton are associated with higher Fock states such as $\ket{uud Q \bar Q}$ in the hadronic eigenstate of the LF Hamiltonian; this implies that the heavy quarks are multi-connected to the valence quarks. The probability for the heavy-quark Fock states scales as $1/m^2_Q$ in non-Abelian QCD.  Since the LF wavefunction is maximal at minimum off-shell invariant mass; i.e., at equal rapidity, the intrinsic heavy quarks carry large momentum fraction $x_Q$.  A key characteristic is different momentum and spin distributions for the intrinsic $Q$ and $\bar Q$ in the nucleon, as manifested, for example, in charm-anticharm asymmetry~\cite{Brodsky:1996hc, Melnitchouk:1997ig}, since the comoving quarks can react differently  to the global quantum numbers of the nucleon~\cite{Brodsky:2015fna}.

IC was also proposed in meson-baryon fluctuation models~\cite{Navarra:1995rq,Pumplin:2005yf}. The possible direct and indirect relevance of  IC in several physical processes has led to many phenomenological calculations involving the existence of a non-zero IC   to explain anomalies in the experimental data and possible signatures of IC in upcoming experiments~\cite{Brodsky:2015fna}. Unfortunately, the normalization of the $\ket{uudc\bar{c}}$ intrinsic charm Fock component in the light-front wavefunctions (LFWF) is unknown. Also, the probability to find a two-body state $\bar{D}^0(u\bar{c})\Lambda^+_c(udc)$ in the proton within the meson-baryon fluctuation models cannot be determined without additional assumptions:  precise constraints from future experiments and/or first-principles calculations are required. 

 The effect of whether the IC parton distribution is either included or excluded in the determinations of charm parton distribution functions (PDFs) can induce changes in other parton distributions through the momentum sum rule, which can indirectly affect the analyses of various physical processes that depend on the input of various PDFs. An estimate of intrinsic charm ($c$) and anticharm ($\bar{c}$) distributions can provide important  information to the understanding of charm quark production  in the EMC experiment~\cite{Aubert:1982tt}. The enhancement of charm distribution in the measurement of the charm quark structure function $F_2^c$ compared to the expectation from the gluon splitting mechanism in the EMC experimental data has been interpreted as evidence for  nonzero IC in several calculations~\cite{Brodsky:1984nx,Harris:1995jx,Hoffmann:1983ah,Brodsky:1991dj}. A precise determination of $c$ and $\bar{c}$ PDFs by considering both the perturbative and nonperturbative contributions is important in understanding charmonia and open charm productions, such as the $J/\psi$ production at large momentum from $pA$ collisions at CERN~\cite{Badier:1983dg}, from $\pi A$ collisions at FNAL~\cite{Leitch:1999ea}, from $pp$ collisions at LHC~\cite{Aaij:2018ogq}, and charmed hadron or jet production from $pp$ collisions at ISR, FNAL, and LHC~\cite{Aaij:2018ogq,Chauvat:1987kb,Aitala:2000rd,Aitala:2002uz}.   LHC measurements associated with cross section of inclusive production of Higgs, $Z$, $W$ bosons via gluon-gluon fusion, and productions of charm jet and $Z^0$~\cite{Aad:2016naf,Aad:2014xaa,Aad:2015auj,Khachatryan:2015oaa},  $J/\psi$ and $D^0$ mesons at  LHCb experiment~\cite{Aaij:2018ogq} can also be sensitive to the  IC distribution. The $J/\psi$ photo- or electro-productions near the charm threshold is believed  to be sensitive to the trace anomaly component  of the proton mass, and some experiments have been proposed at JLab~\cite{E12-12-006} as well as  for the future EIC to measure the production cross section near the threshold. The existence of IC in the proton will provide additional production channels and thus enhance the cross section, especially near the threshold. Similarly, open charm production will also be enhanced by IC. If $c$ and $\bar{c}$ quarks have different distributions in the proton, the enhancements on $D$ and $\bar{D}$ productions will appear at slightly different kinematics. IC has also been proposed to have an impact on estimating the astrophysical neutrino flux observed at the IceCube experiment~\cite{Laha:2016dri}.


In global analyses of PDFs there are different approaches to deal with heavy quarks in which a transition of the number of active quark flavors is made at some scale around the charm quark mass $\mu_c\sim m_c$~\cite{Aivazis:1993pi,Buza:1996wv,Forte:2010ta,Thorne:2012az}.
The transition scale defines where the extrinsic charm-anticharm sea enters. 
However, the intrinsic charm-anticharm sea can exist even at a  lower scale.
In many global fits~\cite{Alekhin:2013nda,Harland-Lang:2014zoa,Ball:2014uwa,Dulat:2015mca,Abramowicz:2015mha,Accardi:2016qay}, the charm quark PDF is set to zero at $\mu_c$, but this is an assumption of no IC. 
In recent years, several PDF analyses started to investigate the possibility of nonzero charm and anticharm distributions at the scale $\mu_c$~\cite{Dulat:2013hea,Pumplin:2007wg,Jimenez-Delgado:2014zga,Ball:2016neh,Hou:2017khm},
but none of them can provide conclusive evidence or exclusion for the intrinsic charm due to the absence of precise data.
A nonzero charm quark PDF at $\mu_c$ is not necessarily  evidence of intrinsic charm, because such estimation depends on the heavy-quark scheme and the $\mu_c$ value used in the fit.
This explains the speculation in~\cite{Dulat:2013hea} that the estimation of intrinsic charm may strongly depend on the choice of the transition scale $\mu_c$.
Fortunately, there is an ideal quantity, the asymmetric charm-anticharm distribution $[c(x)-\bar{c}(x)]$, which would be a clear signal for IC.
Such asymmetry is allowed in QCD because the nucleon has nonzero quark number, the number of quarks minus the number of antiquarks, and thus the $c$ quark and the $\bar{c}$ quark in a nucleon would ``feel'' different interactions, leading to an asymmetric charm-anticharm distribution.
Although the absence of such asymmetry does not exclude intrinsic charm, a nonzero $[c(x)-\bar{c}(x)]$ can serve as  strong evidence, because the extrinsic part of such asymmetry arising at the next-to-next-to-leading order level is negligible~\cite{Catani:2004nc}.

Although the global fits~\cite{Dulat:2013hea,Pumplin:2007wg,Jimenez-Delgado:2014zga,Ball:2016neh,Hou:2017khm} consider the possibility of IC,  all these fits assume $[c(x)-\bar{c}(x)]=0$; constraints on  $[c(x)-\bar{c}(x)]$ have been warranted in the global fit~\cite{Ball:2016neh}. It was found in~\cite{Ball:2016neh}  that a precise and accurate parametrization of the charm PDFs will be useful  for more reliable phenomenology using the data from LHC experiments and will eliminate possible sources of bias arising from the assumptions of only perturbatively-generated charm PDFs. It is therefore  important to determine  if  $[c(x)-\bar{c}(x)]\neq 0$, and how or whether the IC will have significant effect in the physical processes~\cite{Badier:1983dg,Leitch:1999ea,Aaij:2018ogq,Chauvat:1987kb,Aitala:2000rd,Aitala:2002uz,Aad:2016naf,Aad:2014xaa,Aad:2015auj,Khachatryan:2015oaa}. A precise and accurate knowledge of the IC will also have a direct impact on determining the unknown normalization constants of different model calculations associated with the nonperturbative $c(x)$ and $\bar{c}(x)$ distributions. 


An important question to ask is whether the first-principles lattice QCD (LQCD) calculation can provide some constraints or complementary information regarding the existence of  IC. Recently,  there have been LQCD  calculations of the strange $(s)$ quark electromagnetic form factors~\cite{Green:2015wqa,Sufian:2016pex,Sufian:2016vso,Sufian:2017osl,Djukanovic:2019jtp,Alexandrou:2019olr}  with higher precision and accuracy than previously  attained by experiments. These LQCD calculations provided indirect evidence for a nonzero strange quark-antiquark asymmetry in the nucleon by pinning down the nonzero value  of the strange electric form factor $G^s_{E}(Q^2)$ at $Q^2>0$.  The determination of  $G^s_{E,M}(Q^2)$ from the LQCD calculation in~\cite{Sufian:2016pex} has led to precise determination of neutral current weak  axial and electromagnetic form factors~\cite{Sufian:2017osl,Sufian:2018qtw}.  Using LQCD results from~\cite{Sufian:2016pex,Sufian:2016vso} as constraints, it has been shown recently in~\cite{Sufian:2018cpj}, within the  light-front holographic QCD (LFHQCD)  approach~\cite{Brodsky:2006uqa,deTeramond:2008ht,Brodsky:2014yha,Zou:2018eam} and the generalized Veneziano model~\cite{Veneziano:1968yb,Ademollo:1969wd,Landshoff:1970ce}, that the $[s(x)-\bar{s}(x)]$ distribution is  negative  at small-$x$ and positive at large-$x$. This shows the possibility of applying  LQCD results for the phenomenological study of the $[s(x)-\bar{s}(x)]$ asymmetry in the absence of precise experimental data and global fits of the strange quark PDFs. 

The main goal of this article is to calculate the charm electric and magnetic form factors, $G^c_E(Q^2)$ and $G^c_M(Q^2)$,   from LQCD  at nonzero momentum transfer and  discuss their  connection to the existence of IC and a nonzero $[c(x)-\bar{c}(x)]$ asymmetry distribution in the nucleon.  Using the LQCD calculation of $G^c_{E,M}(Q^2)$ as  constraint,   we determine the $[c(x)-\bar{c}(x)]$ distribution using the nonperturbative framework described in~\cite{deTeramond:2018ecg}. We note that the electromagnetic current is odd under charge conjugation and the Dirac form factor $F_1^c(Q^2>0)$ provides a measure of the $c$-quark minus the $\bar{c}$-quark contribution due to the opposite charges of the quark and antiquark. While $F_1^c( Q^2 = 0)=0$,  required by the quantum numbers of the nucleon, a positive $F_1^c(Q^2)$ at $Q^2>0$ implies that the $c$-quark distribution is more centralized  than the $\bar{c}$ quark distribution in coordinate space.  This, in turn, results in a $[c(x)-\bar{c}(x)]$ asymmetry in  momentum space, thereby providing possible evidence for the nonperturbative IC in the nucleon.  On the other hand, a nonzero charm  Pauli form factor $F^c_2(Q^2)$ and a nonzero charm magnetic moment $\mu^c_M\neq 0$ are consequences of  a nonzero orbital angular momentum contribution to the nucleon from charm quarks~\cite{Brodsky:1980zm}. This can be understood in the inherently relativistic LF formalism where a nonzero anomalous magnetic moment requires to have orbital angular momentum $L^z=0$ and $L^z=1$ Fock-states components in the LFWF. 

The  remainder  of  this  article  is  organized  as  follows: In  Section~\ref{sec:lattice},  we  briefly  discuss  the LQCD calculation of the charm electromagnetic form factors $G^c_{E,M}(Q^2)$ and determine these  in the physical limit. In Section~\ref{sec:LFHQCD}, we  use  LQCD results for $G^c_{E,M}(Q^2)$ as input for  quantitative analysis of the $[c(x)-\bar{c}(x)]$ asymmetry distribution in the nucleon within the specific framework of LFHQCD and the generalized Veneziano model. We also present a brief qualitative discussion of our results  in Section~\ref{sec:conclusion}.
 

\section{Lattice QCD calculation of $G^c_{E,M}(Q^2)$} \label{sec:lattice}

 We present in this section the first  lattice QCD calculation of the charm quark electromagnetic form factors in the nucleon. This  first-principles analysis  requires a disconnected insertion  calculation. By ``disconnected insertion," one refers to the nucleon matrix elements involving self-contracted quark graphs (loops), which are correlated with the valence quarks in the nucleon propagator by the fluctuating background gauge fields. ({\it Notice the distinction with the  term ``disconnected diagram" used in the continuum Quantum Field Theory literature.}) Numerical expense and complexity of the disconnected insertion calculations in LQCD, the deficit of  good signal-to-noise ratio in the matrix elements, and  the possibility for a very small magnitude of the $c$ quark matrix elements make it difficult to obtain a precise determination of $G^c_{E,M}(Q^2)$.  We, therefore, need to accept several limitations while performing this calculation. For example, the data is almost twice as noisy compared to the  matrix elements of the strange electromagnetic form factors $G^s_{E,M}(Q^2)$~\cite{Sufian:2016pex,Sufian:2016vso,Sufian:2017osl} and we do not see any signal for one of the gauge ensembles (32ID with a lattice spacing of $a=0.143$ fm~\cite{Blum:2014tka}) used in the previous calculations~\cite{Sufian:2016pex,Sufian:2017osl}. Moreover, we are only able to perform the widely used two-states summed ratio fit of the nucleon three-point ($3pt$) to two-point ($2pt$) correlation functions instead of a simultaneous fit to the summed ratio and conventional $3pt/2pt$-ratio as was done in~\cite{Sufian:2017osl}. The reason is that the $3pt/2pt$-ratio fit for extracting $G^c_{E,M}(Q^2)$ is not stable for the ensemble at the physical pion mass $m_\pi=139$ MeV. We also keep in mind that the $\mathcal{O}(m_c^2a^2)$ errors associated with the lattice spacing can be larger than the case for the $G^s_{E,M}(Q^2)$ matrix elements. 

Our calculation comprises numerical computation with valence overlap fermion on three RBC/UKQCD domain-wall fermion gauge configurations~\cite{Blum:2014tka,Aoki:2010dy}:  (ensemble ID, $L^3\times T$, $\beta$, $a({\rm fm})$, $m_\pi ({\rm MeV})$, $N_{\rm config}$)=\{(48I, $48^3\times 96$, 2.13, 0.1141(2), 139, 81), (32I, $32^3\times 64$, 2.35, 0.0828(3), 300, 309), (24I, $24^3\times 64$, 2.13, 0.1105(3), 330, 203)\}. Here $L$ is spatial and $T$ is temporal size, $a$ is lattice spacing, $m_\pi$ is the pion mass corresponding to the degenerate light-sea quark mass and $N_{\rm config}$ is the number of configurations. We use 17 valence quark masses across these ensembles to explore the quark-mass dependence of the charm electromagnetic form factors. The details of the numerical setup of this calculation can be found in~\cite{Li:2010pw,Gong:2013vja,Yang:2015uis,Sufian:2017osl}. $m_c$ was determined  in a global fit on the lattice ensembles with $\beta=2.13$ fm and $2.25$ using inputs from three physical quantities, such as $M_{D^*_s}$, $M_{D^*_s}-M_{D_s}$, and $M_{J/\psi}$ in~\cite{Yang:2014sea}.  Our statistics are from approximately 100k to 500k measurements across the 24I to 48I ensembles.  The quark loop is calculated with the exact low eigenmodes (low-mode average) while the high modes are estimated with 8 sets of $Z_4$ noise~\cite{Dong:1993pk} on the same $(4,4,4,2)$ grid with odd-even dilution and additional dilution in time. We refer the readers to  previous work~\cite{Sufian:2017osl} for a detailed discussion of the similar numerical techniques which have been used for this calculation.
\begin{figure}[htp]
\centering
\includegraphics[width=2.9in, height=3.8in]{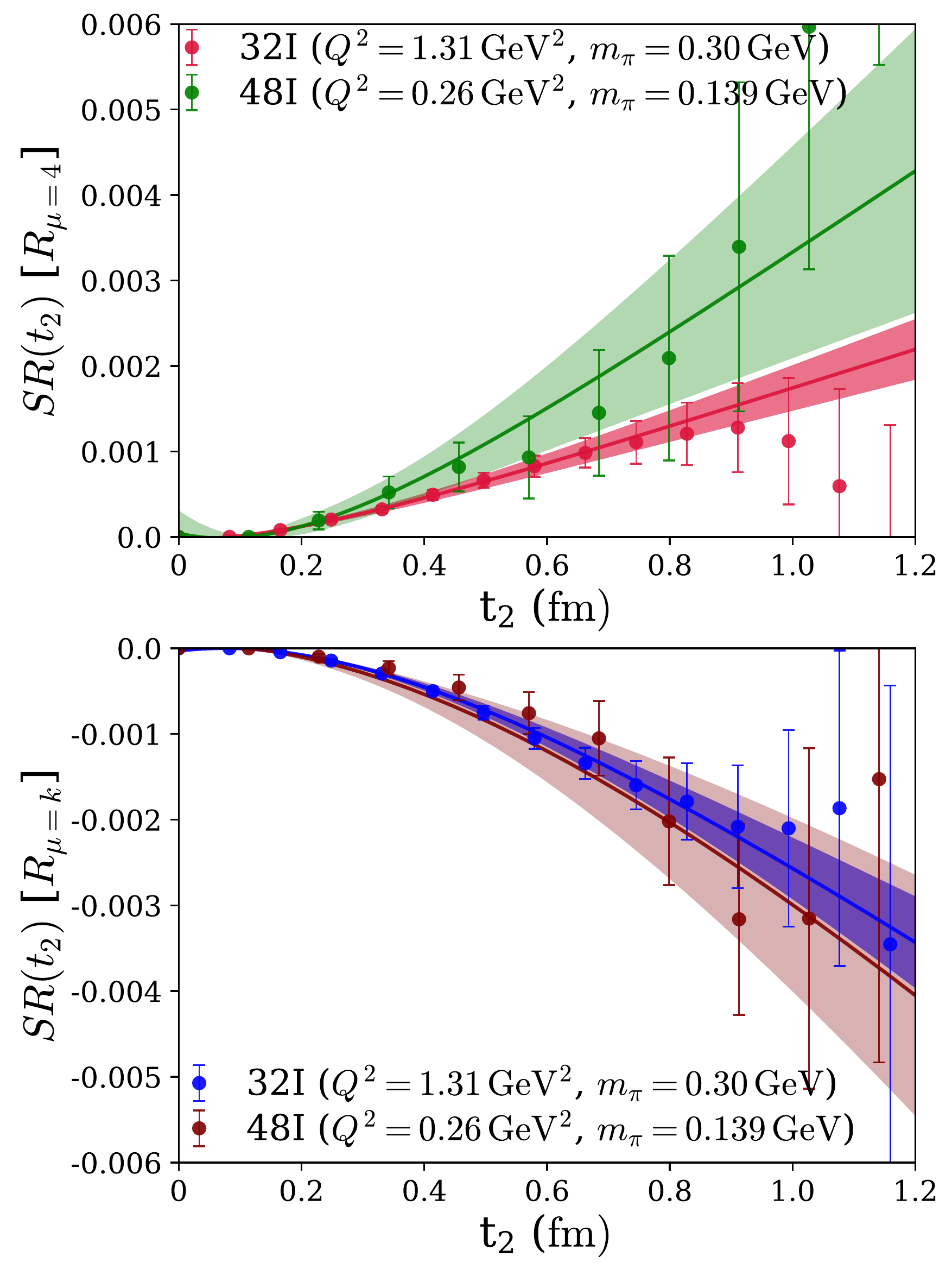}
\caption{ Two-state fits of the 32I and 48I ensembles 3pt/2pt summed ratio data for $G^c_{E,M}(Q^2)$ matrix elements at the unitary points. The colored bands show the fit results. The upper panel shows the fit to matrix elements for $G^c_{E}(Q^2)$ and the lower panel shows that for $G^c_{M}(Q^2)$.\label{fig:sumplot}}
\end{figure}
%
 $G^c_{E,M}(Q^2)$ can be obtained by the ratio of a combination of 3pt and 2pt correlations as,
\bea \label{RatioEq}
R_\mu(\vec{q},t_2, t_1) \equiv &&\frac{\Tr [\Gamma_m \Pi^{3pt}_{V_\mu} (\vec{q},t_2, \!t_1 ) ]} {\Tr [\Gamma_e \Pi^{2pt} (\vec{0},t_2)]} \nn \\
&& \times e^{(E_q-m)\cdot(t_2-t_1)} \frac{2E_q}{E_q+m_N}.
\eea
Here, $\Pi^{2pt}$ is the nucleon 2pt function, $\Pi^{3pt}_{V_\mu}$ is the nucleon 3pt function with the  bilinear  operator $V_\mu(x)= c(x)\gamma_\mu \bar{c}(x)$, $E_q=\sqrt{m_N^2+\vec{q}\,^2}$ and $m_N$ is the nucleon mass, $\vec{q}=\vec{p}\,'-\vec{p}$ is the three-momentum transfer with sink momentum $\vec{p}\,'$ and the source momentum $\vec{p}=0$. The projection operator for $G^c_E$ is $\Gamma_m=\Gamma_e\!=\!(1+\gamma_4)/2$ and that for $G^c_M$ is $\Gamma_m \!=\!\Gamma_k\! =\! -i(1+\gamma_4)\gamma_k\gamma_5/2$ with $k\!=\!1,2,3$. $R_{\mu}$ contains a ratio $Z_P(q)/Z_P(0)\neq 1$, where $Z_P(q)$ is the wavefunction overlap for the point sink with momentum $\vert\vec{q}\vert$. As estimated in~\cite{Sufian:2017osl}, the error introduced by neglecting this factor is about $\sim 5\%$ compared to the statistical error $\geq 30\%$ in the matrix elements and  thus it is ignored in this work. We extract $G^c_{E,M}(Q^2)$ matrix elements using the two-states fit of the nucleon $3pt/2pt$ summed ratio $S\!R(t_2)$ for a given $Q^2$ and fixed index in Eq.~\eqref{RatioEq}:
\begin{align} \label{SR-method}
 & S\!R(t_2) \equiv  \sum_{t_1 \geq t'}^{t_1\leq (t_2-t'')} R(t_2,t_1)=(t_2-t' - t^{''} +1)C_0 +\nn \\
&  C_1 \frac{e^{-\Delta m t''}-e^{-\Delta m (t_2-t' + 1)}}{1-e^{-\Delta m }} + C_2 \frac{e^{-\Delta m t'}\!-\!e^{-\Delta m (t_2-\!t'' \!+\! 1)}}{1-e^{-\Delta m }}. 
\end{align}
Here, $R(t_2,t_1)$ is the $3pt/2pt$-ratio, $t_0$ and $t_2$ are the source  and sink temporal positions, respectively, and $t_1$ is the time at which the bilinear operator $\bar{c}(x)\gamma_\mu c(x)$ is inserted, $t'$ and $t^{''}$ are the number of time slices we drop at the source and sink sides, respectively, and we choose $t'=t''=1$. $C_i$ are the spectral weights involving the excited-state contamination.  Ideally, $\Delta m$ is the energy difference between the first excited state and the ground state but in practice, this is an average of the mass difference between the proton and the lowest few excited states.   As shown in~\cite{Sufian:2017osl}, the $3pt/2pt$-ratio data points are almost symmetric between the source and sink within uncertainty and introducing two $\Delta m$ does not change the fit results of $C_0$. The excited states in the $S\!R(t_2)$ fit fall off faster as $e^{-\Delta mt_2}$ compared to the two-states fit case of the $3pt/2pt$-ratio where the excited-state falls off at a slower rate as $e^{-\Delta m(t_2-t_1)}$. These faster-decreasing excited-state effects allow for fitting the matrix elements starting from shorter time extents, as was demonstrated in~\cite{Chang:2018uxx}.  In Fig.~\ref{fig:sumplot}, we present a sample extraction of $G^c_{E,M}$ matrix elements on 48I at $m_\pi=139$ MeV pion mass at $Q^2=0.25$ GeV$^2$ which enables us to demonstrate the extraction of $\Delta m$ with signal-to-noise ratio better than at other $Q^2$ data points on the 48I ensemble. We also present a similar example on the 32I ensemble at $m_\pi=300$ MeV and at the largest $Q^2$ where the excited-state contribution is expected to be the largest. For example, we obtain $\Delta m= 0.48(29)$, $C_0=0.0005(3)$ on the 48I ensemble and $\Delta m= 0.38(20)$, $C_0=0.00018(4)$ on the 32I ensemble in the $G^c_E(Q^2)$ fits.

\begin{figure}[htp]
\centering
\includegraphics[width=3.15in, height=4.8in]{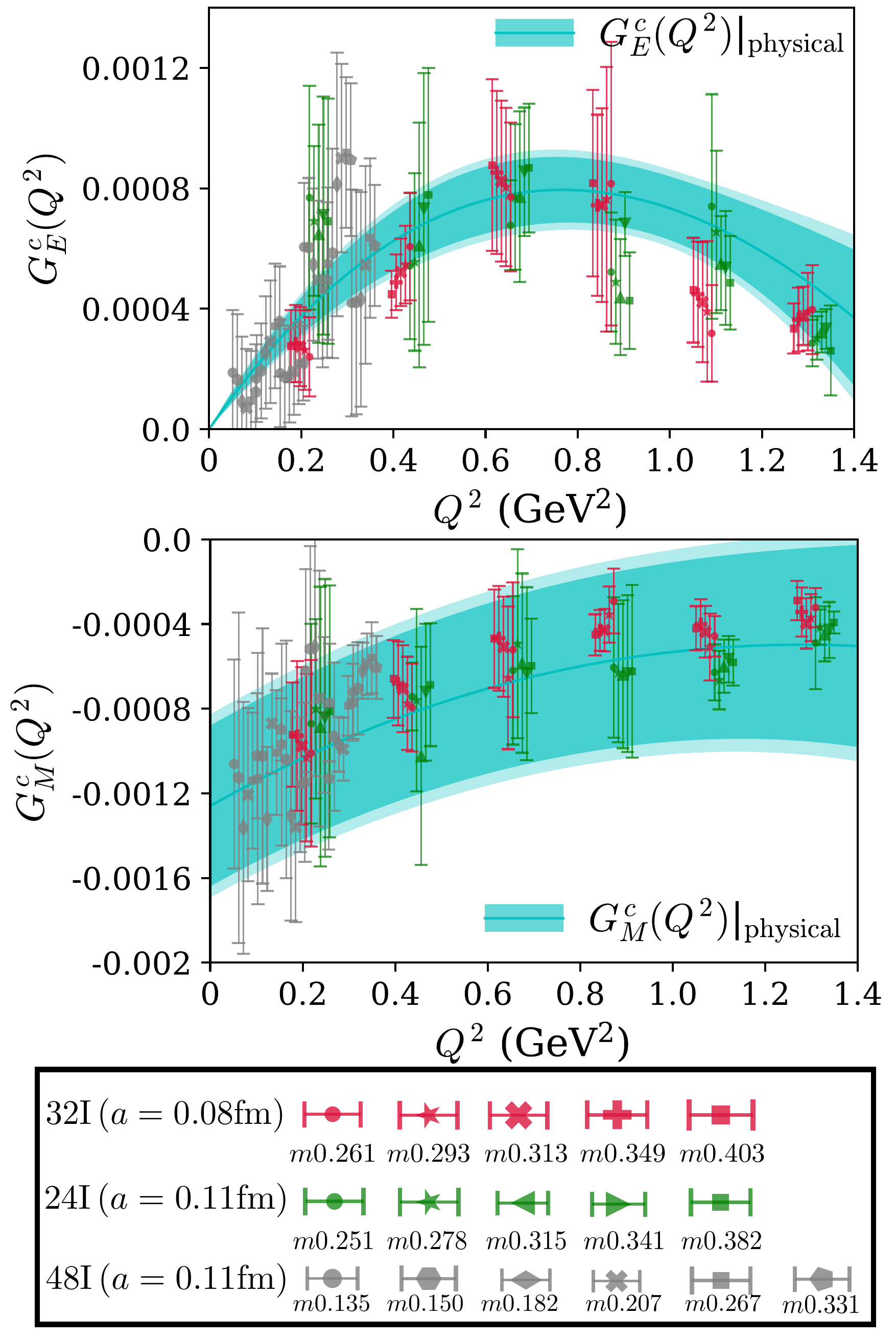}
\caption{$G^c_{E,M}(Q^2)$ matrix elements obtained from the 48I, 32I, and 24I ensembles. Corresponding legends for different pion masses are included in the lower panel of the figure.  The numbers in the legends, such as $m139$, $m251$ represent the data points corresponding to pion mass 139 MeV and 251 MeV, respectively  at different $Q^2$-values. The cyan band indicates  $G^c_{E,M}(Q^2)\vert_{\rm physical}$. The outer (lighter tinted) cyan margins represent an estimate of systematic uncertainty.  Matrix elements at the same $Q^2$-value but at different pion masses are shown with small  offsets  for  better  visibility.\label{fig:GE}}
\end{figure}

We present the matrix elements of $G^c_{E,M}(Q^2)$ obtained from the fit Eq.~\eqref{SR-method}  in the upper and lower panels of Fig.~\ref{fig:GE}.
With the extracted 102 matrix elements from three gauge ensembles (for each of  $G^c_E(Q^2)$ and $G^c_M(Q^2)$) at different pion masses and $Q^2$,  we perform a simultaneous  correlated  and model-independent $z$-expansion fit~\cite{Boyd:1994tt,Bourrely:2008za} to  $G^c_{E,M}(Q^2)$ in the momentum transfer range of $0\leq Q^2 \leq 1.4$ GeV$^2$ and perform chiral, continuum (lattice spacing $a\to 0$), and infinite volume limit (lattice spatial extent $L\to \infty$) extrapolations to obtain the form factors in the physical limit. For such a fit to   $G^c_{E}(Q^2)$, we adopt the following fit form
\begin{align} \label{GEfit}
&G^c_E(Q^2,m_\pi, m_{\pi,vs},m_{J/\psi},a,L ) = \sum_{k=0}^{k_{\rm max}} \lambda_k z^k \times \bigg(1+A_1 m_\pi^2 +\nn \\
& A_2  m_{\pi,vs}^2 + A_3 m_{J/\psi}^2 + A_4 a^2 + A_5 \sqrt{L}\, e^{-m_\pi L}\bigg)\, ,
\end{align}
\bea
{\rm where} \quad z = \frac{\sqrt{t_{\rm cut}+Q^2}-\sqrt{t_{\rm cut}}}{\sqrt{t_{\rm cut}+Q^2}+\sqrt{t_{\rm cut}}}.
\eea
In fit Eq.~\eqref{GEfit}, $m_{\pi, vs}$ is
the partially quenched pion mass {\mbox{$m_{\pi, vs}^2 = 1/2(m_{\pi}^2 + m_{\pi, ss}^2)$}} with $m_{\pi, ss}$ the pion mass corresponding to the sea quark mass. The $m_{J/\psi}$ masses for the lattice ensembles are obtained in~\cite{Yang:2014sea} and extrapolated to the physical value $m_{J/\psi}=3.097$ GeV~\cite{Tanabashi:2018oca}. $A_4$ includes the mixed-action parameter $\Delta_{\rm mix}$~\cite{Lujan:2012wg}. The volume correction in fit~\eqref{GEfit} has been adopted from~\cite{Tiburzi:2014yra} to best describe the data. We use $t_{\rm cut}=m_{J/\psi}^2$, the pole of $c\bar{c}$ pair production. We note that this choice is different from the fit to the strange quark form factor where the $t_{\rm cut}$ is chosen at $4m_K^2$, because the mass of two kaons is less than the mass of $\phi$, while the mass of two $D$ mesons is greater than the mass of $J/\psi$. One may also consider $\eta_c$, which is a bit lighter, but $J/\psi$ is more likely to be produced from a vector current. 

The inclusion of higher-order terms beyond $k_{\rm max}= 4$ has no statistical  significance  and  is  not  considered  in  the $z$-expansion fit~\eqref{GEfit}. We obtain  $\chi^2/{\rm d.o.f.}=1.17$ for the fit~\eqref{GEfit} and the fit parameters are $\lambda_0=0$, $\lambda_1=0.084(15)$, $\lambda_2=-2.38(60)$, $\lambda_3=6.04(9.79)$, $\lambda_4=-0.13(5.79)$, $A_1=-1.05(52)$, $A_2=-0.18(84)$, $A_3=0.025(86)$, $A_4=-0.24(60)$, $A_5=-0.02(34)$.
 Replacing the correction term $A_1 m_\pi^2$ by $A_1 m_\pi$ results in negligible change in the final result. A faster decreasing volume correction $\exp(-m_{D_0}L)$ correction gives $A_5 = 0.008(21)$ which is a smaller correction compared to $\exp(-m_\pi L)$ as expected and they are in statistical agreement. The significant increase of the uncertainty in the physical value of $G^c_{E}(Q^2)$ at larger  $Q^2$ is due to the fact that the data points on the 24I and 32I ensembles are at much heavier pion mass compared to the matrix element at the physical $m_\pi=139$ MeV on the 48I ensemble and there exist no LQCD data points at $Q^2\geq 0.31$ GeV$^2$ on the 48I ensemble. We also see a similar feature for  $G^c_M(Q^2)$ shown in the lower panel of Fig.~\ref{fig:GE}. The cyan band in Fig.~\ref{fig:GE} represents $G^c_{E}(Q^2)\vert_{\rm physical}$ in the physical limit after the quark mass, finite lattice spacing and volume corrections have been implemented using the fit parameters  listed above.  Since most of the $A_i$ corrections do not have statistical significance, we explore the above fit with separate combinations of $A_i$, for example, with $A_1 \& A_4$, $A_1 \& A_5$, and $A_1, A_4, \& A_5$ correction terms. For these fits, we obtain $\{A_1,A_4\}=\{-0.77(18),-0.23(37)\}$, $\{A_1,A_5\}=\{-0.89(22),-0.28(36)\}$, and $\{A_1,A_4,A_5\}=\{-0.86(22),-0.26(37),-0.24(35)\}$, while the physical $G^c_E(Q^2)$ remains essentially unchanged with slightly smaller final uncertainties compared to when all $A_i$ corrections are included. A similar investigation for the $G^c_{M}(Q^2)$ fit results in a similar conclusion.   

The systematic uncertainty is estimated by calculating the differences between  $G^c_E(Q^2)\vert_{\rm physical}$ and $G^c_E(Q^2)$ obtained from the fit Eq.~\eqref{GEfit} by considering the corrections of the $A_3$, $A_4$, $A_5$ terms from the $m_{J/\psi}$-value on the 24I ensemble obtained in~\cite{Yang:2014sea}, the smallest lattice spacing from the 32I ensemble, and the 48I ensemble with the largest volume, respectively.  The systematic uncertainty has been added as lighter-tinted margins to the statistical uncertainty band in Fig.~\ref{fig:GE}. 

To obtain  $G^c_{M}(Q^2)$ in the physical limit and in the $0\leq Q^2 \leq 1.4$ GeV$^2$ momentum transfer region, we adopt the following empirical fit form with the volume correction term adopted from~\cite{Beane:2004tw}:
\begin{align} \label{GMfit}
& G^c_M(Q^2,m_\pi, m_{\pi,vs},m_{J/\psi},a,L ) = \sum_{k=0}^{k_{\rm max}} \lambda_k z^k \times \bigg(1+A_1 m_\pi^2+ \nn \\
&A_2  m_{\pi,vs}^2 + A_3 m_{J/\psi}^2 + A_4 a^2 + A_5 m_\pi\, \bigg[1-\frac{2}{m_\pi L}\bigg]e^{-m_\pi L}\bigg).
\end{align}
We limit the $k_{\rm max}=3$ in our fit. Additional terms in the $z$-expansion have no statistically significant effect on  $G^c_M(Q^2)\vert_{\rm physical}$. With the $\chi^2/{\rm d.o.f.}=1.14$ in the fit~\eqref{GMfit}, we obtain the fit parameters $\lambda_0=-0.00127(38)$, $\lambda_1=0.054(20)$, $\lambda_2=-0.86(70)$, $\lambda_3=0.45(8.22)$, $A_1=0.0007(25)$, $A_2=-0.001(4)$, $A_3=0.0002(6)$, $A_4=0.001(4)$, and $A_5=-0.029(40)$.
The systematic uncertainty of $G^c_M(Q^2)\vert_{\rm physical}$ is obtained in a similar way  as for the case of $G^c_E(Q^2)\vert_{\rm physical}$.  

The electric and magnetic radii can be extracted from the slope of the $G^c_{E,M}(Q^2)$ form  factors as $Q^2\to 0$:
\bea
\langle r^2_{E,M}\rangle^c = -\,6\,\frac{d G^c_{E,M}(Q^2)}{dQ^2}\bigg\vert_{Q^2=0} = -\,6\,\bigg(\frac{\lambda_1}{4t_{\rm cut}}\bigg) .
\eea
Using the $\lambda_1$-values from above, we obtain
\bea
\langle r^2_{E}\rangle^c &=& -\,0.0005(1)\,\, {\rm fm}^2 ,\nn \\
\langle r^2_{M}\rangle^c &=&  -\,0.0003(1)\,\, {\rm fm}^2. 
\eea

While the existence of a nonzero $G^c_E(Q^2)$ (and thus  the Dirac $F_1^c(Q^2)$ form factor) is related to a nonzero asymmetry of the $[c(x)-\bar{c}(x)]$ distribution, a nonzero $G^c_M(Q^2)$ (and thus Pauli $F_2^c(Q^2)$ form factor) immediately implies that there is a nonzero orbital angular momentum contribution to the nucleon from the charm quarks. The Pauli form factor $F_2^c(Q^2)$ has the LFWF representation as the overlap of the states differing by one unit of orbital angular momentum. It is thus closely related  to the spin sector of the charm quark sea in the proton. $F_2^c(Q^2)$ is also given by the first $x$ moment of the generalized parton distribution $E^c(x,\xi,t)$, $t = - Q^2$, which contributes to the second term of Ji's sum rule~\cite{Ji:1996ek}.  Therefore, our result $\mu_M^c=-0.00127(38)_{\rm stat}(5)_{\rm sys}$ indicates  a nontrivial role of the charm quark sea in understanding the spin content of the proton.

\section{A nonperturbative model for computing the intrinsic $[c(x)-\bar{c}(x)$]  asymmetry in the nucleon \label{sec:LFHQCD}}
 Although a model-independent determination of IC distributions from the charm quark form factors is  not possible at the moment, the underlying physics, governed by QCD, does imply  some  important connections  and constraints, such as the QCD inclusive-exclusive connection~\cite{Drell:1969km,Blankenbecler:1974tm,Brodsky:1979qm}: It relates hadron form factors at large $Q^2$ to the hadron structure function at $x \to 1$, leading to simple counting rules.  Furthermore, since the $c$ and $\bar{c}$ carry opposite charges, the charm quark form factor measures the difference of the transverse charge density~\cite{Miller:2018ybm} between the $c$ and $\bar{c}$ quarks. For a positive charge form factor, like those shown in Fig.~\ref{fig:GE}, the charm quark distribution is more spread out than the anticharm distribution in  ${\bf q}_T$-space, where ${\bf q}_T$ is the Fourier conjugate variable of the transverse coordinate ${\bf b}_T$ and ${\bf q}_T^2=Q^2$. As a feature of Fourier transform, the charm quark density $\rho({\bf b}_T)$ is more centralized in ${\bf b}_T$-space than the anticharm quark. Representing the density as the square of the wave function, $\rho({\bf b}_T)=|\tilde{\psi}({\bf b}_T)|^2$, one can easily find the ${\bf k}_T$-space wave function $\psi({\bf k}_T)$ is more spread out for the charm quark, where ${\bf k}_T$ is the intrinsic transverse momentum. Thus the actual distribution for a positive charge form factor favors the $c$ quark  carrying higher momentum than the $\bar{c}$ quark. As a result, the $[c(x)-\bar{c}(x)]$ distribution will favor negative values in the low-$x$ region and positive values in the high-$x$ region. We note that a strict definition of the transverse charge density is given by the Fourier transform of the Dirac form factor $F_1^c(Q^2)$, which dominates the charge form factor $G_E^c(Q^2)$ in the low-$Q^2$ regime.

For a quantitative estimation of  $[c(x)-\bar{c}(x)]$, one currently has to rely on additional assumptions, although the form factor result does indicate some qualitative features of the distribution function based on  the discussions above. Here we take the  nonperturbative  phenomenological model in~\cite{Sufian:2018cpj}, which relates the form factor and sea-quark  distribution functions with minimal parameters. The formalism is  based on the gauge/gravity correspondence~\cite{Maldacena:1997re}, light-front holographic mapping~\cite{Brodsky:2014yha,Zou:2018eam,deTeramond:2018ecg}, and the generalized Veneziano model~\cite{Veneziano:1968yb,Ademollo:1969wd,Landshoff:1970ce}. In the following, we refer to this model  as LFHQCD. The charm quark Dirac and Pauli form factors are given by~\cite{Sufian:2016hwn}
\bea
    F_1^c(Q^2) &=& \sum_{\tau} c_\tau [F_{\tau}(Q^2) - F_{\tau+1}(Q^2)],\label{eq:F1}\\
    F_2^c(Q^2) &=& \sum_{\tau}\chi_{\tau} F_{\tau+1}(Q^2),\label{eq:F2}
\eea
 where $\tau$ is the number of constituents of the Fock state component. The leading Fock state with nontrivial contribution to the charm form factor is $|uudc\bar{c}\rangle$, which is a $\tau=5$ state. With additional intrinsic sea quark pairs, the Fock states, {\it e.g.} $|uudu\bar{u}c\bar{c}\rangle$, $|uudd\bar{d}c\bar{c}\rangle$, etc., will contribute to $\tau=7$ terms. If also considering the possibility of intrinsic gluon constituents, one may have the contribution from $|uudc\bar{c}g\rangle$, $\tau=6$, and/or higher Fock states.

Form factor $F_{\tau}$ can be expressed in a reparametrization invariant form~\cite{deTeramond:2018ecg} 
 \begin{align} \label{eq:FFw}
 F(t)_\tau = \frac{1}{N_\tau} \int_0^1 dx\, w'(x) w(x)^{-\alpha(t)} \left[1 - w(x) \right]^{\tau -2} ,
\end{align}
where $\alpha(t)$  is the Regge trajectory, and $N_\tau$ is a normalization factor; $w(x)$ is a flavor independent function with  $w(0) = 0, \, w(1) = 1$ and   $w'(x) \ge 0$. We use the same  universal form of the function $w(x)$ from~\cite{deTeramond:2018ecg}
\bea
    w(x) = x^{1-x} e^{-a (1-x)^2},
\eea 
with $a=0.480$~\cite{Liu:2019vsn}: It incorporates Regge behavior at small $x$ with the $J/\psi$ intercept~\eqref{Jpsi}, $w(x) \to x$ as $x \to 0$, and the inclusive-exclusive counting rule at large $x$, $q_\tau(x) \to (1 - x)^{2 \tau -3}$, as $x \to 1$~\cite{deTeramond:2018ecg}. The light front holographic approach leading to these results is based on the underlying conformal algebra which leads to linear Regge trajectories.  The spin-flavor coefficients $c_\tau$ and $\chi_\tau$ are parameters to be  determined from the LQCD computation of $G^c_E(Q^2)\vert_{\rm physical}$ and $G^c_M(Q^2)\vert_{\rm physical}$ to obtain $F_1^c(Q^2)$. The constraint that the numbers of charm and anticharm quarks are identical for each Fock state component has been  incorporated  in Eq.~\eqref{eq:F1}.

Then the asymmetric charm-anticharm distribution function is
\bea
    c(x) - \bar{c}(x) = \sum_\tau c_\tau [q_\tau(x) - q_{\tau+1}(x)],\label{eq:ccbar}
\eea
where $\tau \ge 5$ and
\bea \label{eq:qtau}
q_\tau(x) = \frac{1}{N_\tau} w(x)^{-\alpha(0)} [1-w(x)]^{\tau-2} w'(x).
\eea
We should note here that the coefficients $c_\tau$ in  Eq.~\eqref{eq:ccbar} are the same as those in Eq.~\eqref{eq:F1}. Therefore, once they are determined by the form factor, one can make predictions for the distribution  functions.

The form factors and distribution functions above are derived at the massless quark limit and one may have different approaches to incorporate quark mass corrections. For small quark masses (up, down and strange) the latter can be treated perturbatively, leaving the Regge slope unchanged and leading to a moderate change of the intercept. The resulting spectra are in very good agreement with experiment~\cite{Brodsky:2014yha,Brodsky:2016yod}. The situation is more intricate  for the case of  heavy quarks, like  $c$ quarks, since now conformal symmetry is strongly broken and the occurrence  of linear trajectories is far from obvious. It has been shown, however, that the formalism can  indeed be extended  to heavy quark bound states~\cite{Dosch:2016zdv,Nielsen:2018ytt}, leading to a fair  agreement with the data. In  this case, the Regge trajectories are still linear, but the  slope depends on the heavy quark mass. The intercept changes quite drastically with the quark mass.

The $J/\psi$ Regge trajectory obtained in~\cite{Nielsen:2018ytt} is
\bea \label{Jpsi}
\alpha(t)_{J/\psi}= \frac{t}{4\kappa_c^2} - 2.066,
\eea
where $\kappa_c=0.874\,\rm GeV$.
This result agrees with the one obtained in a phenomenological potential model~\cite{Chaturvedi:2018xrg}. The large change of the intercept as compared to light quarks removes the small-$x$ singularity of quark distribution functions while keeping the counting rules at large $Q^2$ and at large $x$ unchanged. The change of the slope affects  only the generalized  parton distribution function. 
The quark distribution difference $[c(x)-\bar{c}(x)]$ is not sensitive to the choice of  the mass correction procedure, since the quark mass affects equally charm and anticharm distributions.

In practice, one needs to truncate the expansion in Eq.~\eqref{eq:F1} to have numerical results. For simplicity, we only keep the lowest Fock state containing the charm quark components,  {\it i.e.}, $\tau=5$. 
The coefficient $c_\tau$ is determined, through Eqs.~\eqref{eq:F1} and \eqref{eq:F2} by the lattice results of $G_E^c(Q^2)$ and $G_M^c(Q^2)$ at the physical limit. We perform a fit to the extracted results of $G_E^c(Q^2)\vert_{\rm physical}$ and $G_M^c(Q^2)\vert_{\rm physical}$, {\it i.e.,} the bands in Figs.~\ref{fig:GE}. 
Since the lattice data from different ensembles are evaluated at different $Q^2$ values, and have been utilized to determine the quark mass, lattice spacing, and finite volume effects, the effective number of data points in the physical limit is 6 for $G_E^c(Q^2)|_{\rm physical}$ and 6 for $G_M^c(Q^2)|_{\rm physical}$\footnote{For each ensemble we have data points at 6 different $Q^2$. A simultaneous fit of the data from three ensembles (48I, 32I, 24I) with different quark masses, lattice spacings, and volumes leads to the results in the physical limit.}. To really capture the uncertainty, we create 200 replicas from the extracted bands. Each replica is firstly generated by randomly sampling 6 data points of $G_E^c(Q^2)\vert_{\rm physical}$ and 6 data points of $G_M^c(Q^2)\vert_{\rm physical}$ from the extracted bands within $0<Q^2<1.4\,\rm GeV^2$, which are covered by the lattice data. Then for each data point, the central value is resampled with a Gaussian distribution according to its uncertainty. In addition, we also randomly shift the value of $\kappa_c$ within $\pm5\%$ in each single fit of one replica to incorporate the theoretical uncertainty.  The coefficient determined from the fit is $c_{\tau=5}=0.018(3)$.

\begin{figure}[htp]
\begin{center}
\setlength\belowcaptionskip{-2pt}
\includegraphics[width=3.2in, height=2.3in]{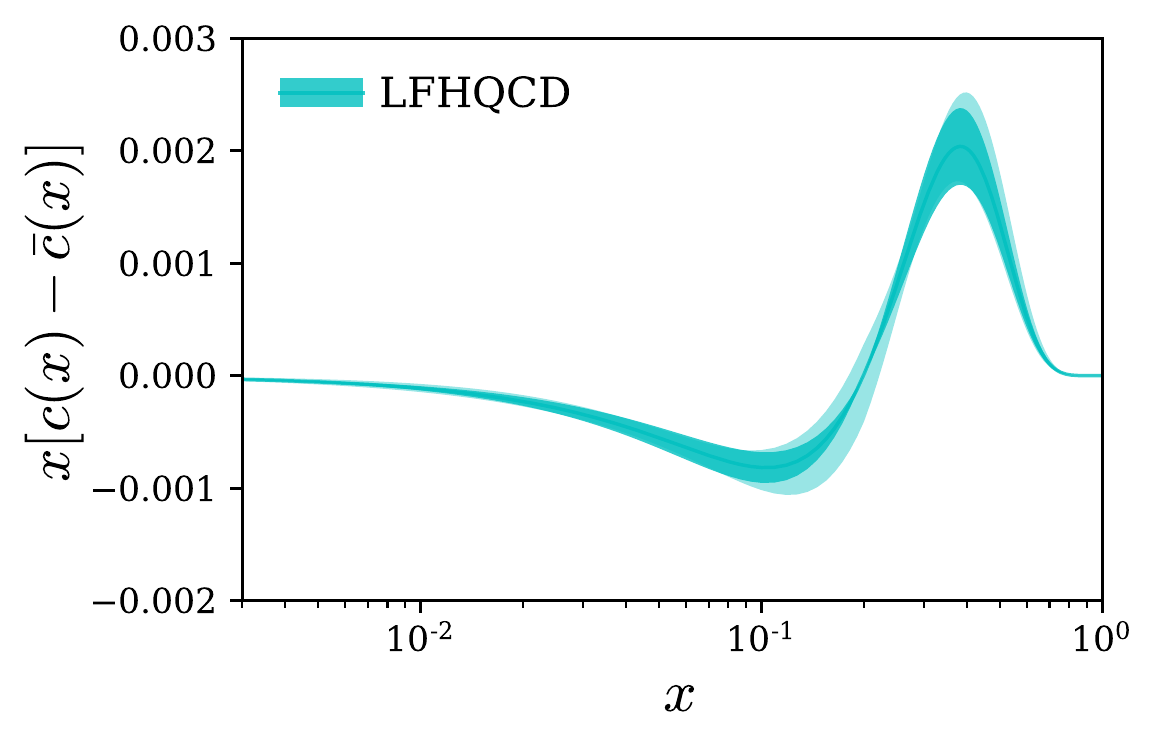}
\caption{The distribution function $x[c(x)-\bar{c}(x)]$ obtained from the LFHQCD formalism using the lattice QCD input of charm electromagnetic form factors $G^c_{E,M}(Q^2)$\label{fig:ccbardis}. The outer (lighter tinted) cyan margins represent an estimate of systematic uncertainty  in the $x[c(x)-\bar{c}(x)]$ distribution obtained from a variation of  the hadron scale $\kappa_c$ by 5\%.}
\end{center}
\end{figure}

 Having obtained the charm coefficient $c_{\tau = 5}$ from the lattice computation, we use  Eq.~\eqref{eq:ccbar}, to  obtain the asymmetric charm-anticharm distribution function $x[c(x)-\bar{c}(x)]$ shown in Fig.~\ref{fig:ccbardis}. The result from the fit is in agreement with the qualitative analysis at the beginning of this section, namely, that the charm quark tends to carry larger momentum than the anticharm quark based  on the lattice results for  the charm quark form factors. 
From the $x[c(x)-\bar{c}(x)]$ distribution obtained by combining LQCD results from  $G^c_{E,M}(Q^2)$ and the LFHQCD formalism, we can calculate the first moment of the difference of $c(x)$ and $\bar{c}(x)$ PDFs to be
\begin{align}
\langle x\rangle_{c-\bar{c}} = \int^1_0 dx\, x\, [c(x)-\bar c(x)]=0.00047(15),
\end{align}
 where the total uncertainty is obtained from the fitting error in $c_{\tau=5}$ and 5\% variation in $\kappa_c$. The $[c(x)-\bar{c}(x)]$ distribution result is about 3 times smaller in magnitude than the $s(x)-\bar{s}(x)$ distribution obtained with the same formalism~\cite{Sufian:2018cpj}. Although a small asymmetry could be a result of the cancellation of two relatively large $c(x)$ and $\bar{c}(x)$ distributions, it is possible that the intrinsic charm and anticharm distributions are both small. 
Furthermore, the charm and anticharm distributions at high energy scales are dominated by the extrinsic sea from perturbative   radiation. 
The experimental observation and isolation of the intrinsic charm effect are extremely challenging in such cases. Thus it is not surprising that the recent measurement of $J/\psi$ and $D^0$ productions by the LHCb collaboration~\cite{Aaij:2018ogq} found no intrinsic charm effect.
An ideal place to investigate intrinsic charm would be the $J/\psi$ or open charm productions at relatively low energies, {\it e.g.}, at JLab, although it is also possible to see intrinsic charm effects in very accurate measurements of high energy reactions. In addition,  lepton-nucleon scattering may provide a cleaner probe than nucleon-nucleon scattering to help reduce backgrounds and increase the chance to observe the intrinsic charm effect, and therefore the future EIC will provide such opportunities.

The nonzero value of $G^c_E(Q^2)$ can also originate from the interference of the $q\rightarrow gq \rightarrow c\bar{c}q$ and $q \rightarrow ggq \rightarrow c\bar{c}q$ sub-processes, without the existence of IC. However, as mentioned earlier, this extrinsic $[c(x)-\bar{c}(x)]$ asymmetry which  arises at the next-to-next-to-leading order level is negligible~\cite{Catani:2004nc}. Moreover, according to~\cite{Catani:2004nc}, this extrinsic asymmetry would result in a much smaller and negative value of the first moment of $[c(x)-\bar{c}(x)]$ distribution $\langle x\rangle_{c-\bar{c}}$ compared to $\langle x\rangle_{c-\bar{c}}=0.00047(15)$ obtained in this calculation. A  negative value for $\langle x\rangle_{c-\bar{c}}$ would also result in a positive $[c(x)-\bar{c}(x)]$ distribution at small $x$ and a negative distribution at large $x$, in contrast to the $[c(x)-\bar{c}(x)]$ distribution we have obtained here. But the evidence based on the $[ s(x) - \bar{s}(x)]$ distribution in~\cite{Sufian:2018cpj}, the EMC measurement~\cite{Aubert:1982tt}, and perturbative QCD computation~\cite{Catani:2004nc} seem to indicate extremely small values of extrinsic charm for $x>0.1$.  The present determination of the  $[c(x) - \bar c(x)]$ distribution from LQCD supports the existence of nonperturbative intrinsic heavy quarks in the nucleon wavefunction at large $x \sim 0.3 - 0.5$ with a magnitude consistent with experimental signals. A consequence of this result is  Higgs production at large $x_F> 0.8$ in $p p $ collisions at the LHC from the direct coupling of the Higgs to the intrinsic heavy quark pair~\cite{Brodsky:2007yz}.

\section{Conclusion and outlook \label{sec:conclusion}}

In this article, we have presented the first lattice QCD calculation of the charm quark electromagnetic form factors in the physical limit.  This first lattice QCD calculation indicates that a nonzero charm electric form factor corresponds to the intrinsic charm-anticharm asymmetry in the nucleon sea, thereby providing an  indication of the existence of nonzero intrinsic charm based on a first-principles calculation. In addition, the nonzero value of the charm magnetic form factor indicates a nonzero orbital angular momentum contribution to the nucleon coming from the charm quarks. We have discussed that the existence of IC is supported by QCD and how an accurate knowledge of the intrinsic charm can help to remove bias in the global fits of PDFs and related phenomenological studies. 

Motivated by the new lattice results, we have used the nonperturbative light-front holographic   framework incorporating the QCD inclusive-exclusive connection at large $x$ to determine the $[c(x)-\bar{c}(x)]$ asymmetry up to a normalization factor, which is constrained by the lattice QCD calculation. Since the LFHQCD calculation starts from a nucleon  Fock state with hidden charm, the parton distributions determined in this model refer exclusively to intrinsic charm where the small-$x$ behavior is determined by the $J/\psi$ intercept. On the other hand, contributions from gluon splitting are supposed to be determined by the pomeron trajectory with a much higher intercept. These features will be discussed in a separate publication. 

The new determination of the $[c(x)-\bar{c}(x)]$  asymmetry presented here gives  additional elements and further insights into the existence of intrinsic charm. It also can provide complementary information to the global fits of PDFs which look for  the possibility of IC in the absence of ample experimental data.

\section*{Acknowledgements}

RSS thanks Jeremy R. Green, Luka Leskovec, Jian-Wei Qiu, Anatoly V. Radyushkin, and David G. Richards for useful discussions. The authors thank the RBC/UKQCD collaborations for providing their DWF gauge configurations. This work is supported by the U.S. Department of Energy, Office of Science, Office of Nuclear Physics under contract DE-AC05-06OR23177. A.~Alexandru is supported in part by U.S. DOE Award Number DE-FG02-95ER40907. T. Draper and K.F. Liu are supported in part by DOE Award Number {\text{DE-SC0013065}}. Y. Yang is supported by Strategic Priority Research Program of Chinese Academy of Sciences, Grant No. XDC01040100.  This research used resources of the Oak Ridge Leadership Computing Facility at the Oak Ridge National Laboratory, which is supported by the Office of Science of the U.S. Department of Energy under Contract No. DE-AC05-00OR22725. This work used Stampede time under the Extreme Science and Engineering Discovery Environment (XSEDE), which is supported by National Science Foundation grant number ACI-1053575. We also thank the National Energy Research Scientific Computing Center (NERSC) for providing HPC resources that have contributed to the research results reported within this paper. We acknowledge the facilities of the USQCD Collaboration used for this research in part, which are funded by the Office of Science of the U.S. Department of Energy. 







\end{document}